# Near-infrared Schottky silicon photodetectors based on two dimensional materials

Teresa Crisci, University of Campania "Luigi Vanvitelli – Dipartement of Mathematics and Physics (DMF), Naples, Italy. Institute of Applied Science and Intelligent Systems (ISASI) – National Research Council (CNR), Naples, Italy.

Luigi Moretti, University of Campania "Luigi Vanvitelli – Dipartement of Mathematics and Physics (DMF), Naples, Italy.

M. Gioffrè, Institute of Applied Science and Intelligent Systems (ISASI) – National Research Council (CNR), Naples, Italy.

M. Casalino, Institute of Applied Science and Intelligent Systems (ISASI) – National Research Council (CNR), Naples, Italy. Corresponding: maurizio.casalino@na.isasi.cnr.it

## Abstract

Since its discovery in 2004, graphene has attracted the interest of the scientific community due to its excellent properties of high carrier mobility, flexibility, strong light-matter interaction and broadband absorption. Despite of its weak light optical absorption and zero band gap, graphene has demonstrated impressive results as active material for optoelectronic devices. This success pushed towards the investigation of new two-dimensional (2D) materials to be employed in a next generation of optoelectronic devices with particular reference to the photodetectors. Indeed, most of 2D materials can be transferred on many substrates, including silicon, opening the path to the development of Schottky junctions to be used for the infrared detection. Although Schottky near-infrared silicon photodetectors based on metals are not a new concept in literature the employment of two-dimensional materials instead of metals is relatively new and it is leading to silicon-based photodetectors with unprecedented performance in the infrared regime.
This chapter aims, first to elucidate the physical effect and the working principles of these devices, then to describe the main structures reported in literature, finally to discuss the most significant results obtained in recent years.

## 1. Introduction

In the last few decades, the enormous evolution of social networks and the progress of the Internet of Things (IoT) has made necessary the management of a huge amount of data. For this reason, industry and scientific research has been focused on the development of new technologies to support and to manage the data traffic increase.

Silicon photonics (SP) fits perfectly into this scenario, since it combines the advantages of the mature silicon technology developed in microelectronics with the possibility to further reduce costs simultaneously increasing the transmission speed thanks to the use of light. Hence, SP is currently emerging as an appealing market promising to reach $4 billion in 2025 (fig. 1) [1]. Nowadays, Intel and Luxtera play leadership roles in the SP industry, bringing products to market that can support 100Gb/s of communication throughput.

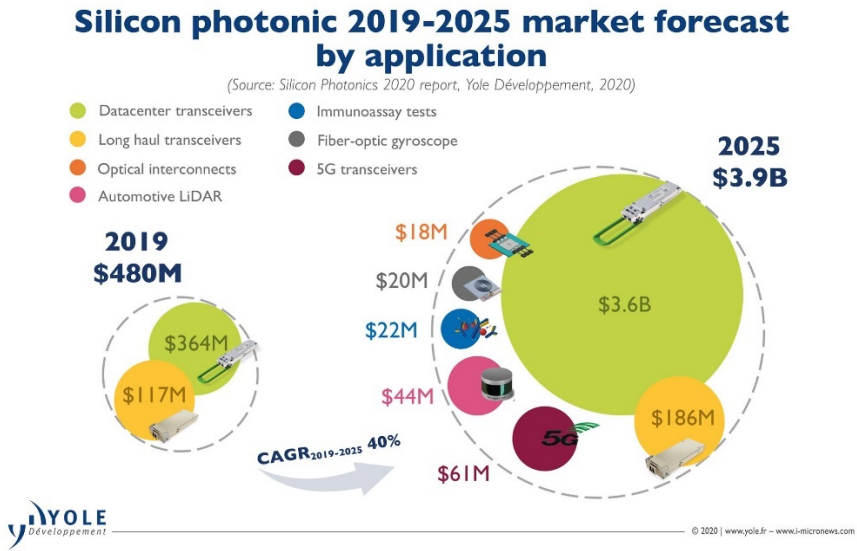

Figure 1. Silicon photonics 2019–2025 market forecast [1].

Silicon photonics is by now a widely consolidated field originating from the pioneering work of Soref et al. [2,3] and from the manufacture of silicon on insulator (SOI) substrates particularly suitable to the realization of guiding structures, both in the 1980s. Thus, the possibility to build optoelectronic devices by using CMOS facilities, would allow not only the low costs advocated by telecommunications industry but also the possibility to integrated both electronic and photonic functionalities on the same chip.

In this context silicon-based photodetectors (Si PDs) are a key component able to establish a connection between the world of electronics and photonics. Si PDs working in the visible spectrum can be easily found on the market, however the telecommunications industry requires components operating in the infrared regime, where, unfortunately, silicon has a negligible absorption due to its indirect bandgap of 1.12 eV.

To overcome this drawback the most common approach is based on the integration of germanium (Ge) on silicon. Nonetheless, the performances of these devices are often limited by a relatively high leakage current caused by the lattice mismatch with silicon of 4,3%. This effect can be mitigated by growing a Ge buffer layer on Si by a two steps epitaxial method giving rise to problems of thermal budget and planarity that limit the monolithic Ge integration on Si.

For all these reasons, an all-silicon approach is preferable, and the exploitation of the absorption phenomena based on the internal photoemission effect (IPE) in a

Schottky diode is among the most promising and innovative.

In a Schottky diode configuration, the photons incident on a Metal-Si interface, with an energy below the silicon bandgap, can cause the generation of photo-excited carriers in the metal with energy higher than Schottky barrier. This "hot" carriers are injected into the silicon, accelerated by the electric field in the depletion region of the junction and collected as a photocurrent. [4-7].

In literature, several examples of IR charged coupled devices (CCDs) based on a Schottky diode can be easily found. The most common example of this family of PDs is based on $Pd_2Si/Si$ and PtSi/Si Schottky junctions used for aerospace applications. The main problem with these photodetectors, however, is the requirement of cryogenic temperatures to minimize the noise current due to the low Schottky barrier (SBH) necessary to achieve a reasonable device efficiency. [8-13].

Consequently, to exploit the IPE at room temperature, new classes of devices characterized by higher values of SBH, have been proposed. Obviously, this approach leads to a worsening of the performances of PDs and, therefore, different solutions have been investigated. Some devices use a Fabry-Perot type resonant geometry for compensating the reduction in efficiency [14,15], others use nanometric metallic structures such as Si nanoparticles (NP) [16], gratings [17] and antennas [18]. Lastly, PDs based on the IPE at room temperature have also been realized by taking advantage of surface plasmonic polaritons on metal strips of nanometric scale (SPP) [19,20]. Despite of these efforts, however, it was possible to obtain a maximum responsivity of only 30 mA/W for PDs integrated in waveguide configuration [16].

To increase these low responsivity values, deriving from the small probability of the photo-excited carriers of overcoming the Schottky barrier, the reduction of the metal thickness has stood out as a good strategy [21,22], influencing the research towards the integration of 2D materials with Si. In particular, 2D layered materials have emerged thanks to their exceptional optical and electronic properties, low cost and simple fabrication process.

In literature it's possible to find various graphene/Si PDs based on FET structures [23,24]; however, such kind of devices suffer of a high dark current needing of interdigitated electrodes because the electric field in graphene is formed in a small region within 200nm from the contact. On the other hand, by taking advantage of an IPE approach, it is possible to minimize the dark current thanks to the rectifying nature of Schottky diodes that don't need of interdigitated structures.

Graphene has opened the way for the investigation of other 2D layered materials. Notable attention has been given to transition-metal dichalcogenides (TMDCs) since their very naturally abundant and possess a tunable bandgap in addition to most of graphene properties [25-27]. Recently, several heterostructures TMDCs/Si have been investigated: the formation of a potential barrier at the interface between the two materials has allowed the exploitation of the IPE to realize high detectivity and ultrafast NIR PDs [28-31].

In this chapter the topic of NIR PDs based on 2D materials/silicon junctions is discussed. In the first section the theoretical background behind the behaviour of junctions based on 2D materials with particular reference to graphene will be explored and compared to the classical theory describing the Schottky junctions using 3D metals. In the second part, several examples of NIR PDs exploiting 2D materials/silicon junctions reported in literature will be presented and discussed.

## 2. Theoretical Background

The responsivity R of a photodetector can be defined as the ratio between the photogenerated current ($I_{ph}$) to the incident optical power ($P_{inc}$). It is very important for the quantification of the PD performance since it is strictly related to the efficiency

of the device. This relation is explicated by the following formula:

$$R = \frac{I_{ph}}{P_{inc}} = \frac{\lambda[nm]}{1242} \cdot \eta_{ext} \qquad (1)$$

where $\eta_{ext}$ is the external quantum efficiency, that represents the number of charged carriers generated for each incident photon. The external quantum efficiency depends on the internal quantum efficiency by the equation $\eta_{ext} = A\,\eta_{int}$, where $\eta_{int}$ is the ratio of the number of charged carriers generated to the number of absorbed photons and A is the active material absorption.

The first theoretical model of photoemission from metal to vacuum was published by Fowler in 1931 [32]. Afterwards, in the 60s, the Fowler's theory was extended to the photoemission in the semiconductor by Cohen, Vims and Archer [33] and Elabd and Kosonocky [21].

By following the Elabd approach, it is possible to obtain the expression of $\eta_{int}$ by starting with the consideration of the number of excited carriers $N_T$ is:

$$N_T = \int_0^{h\nu} D(E)dE \qquad (2)$$

where hv is the incident photon energy, E is the carriers energy referred to Fermi level and the argument function of the integral D(E) is the absorber material density of state (DOS). On the other hand, not all the excited carriers can be emitted from the metal into semiconductor, indeed only those localized to energies higher than Schottky barrier have a certain probability of being emitted. Therefore, the number of states occupied by charge carriers that have a probability P(E) of being emitted in the silicon can be written as:

$$N = \int_{q\phi_{B0}}^{h\nu} D(E)P(E)dE \qquad (3)$$

where P(E) is the charge carrier emission probability.

Elabd and Kosonocky formulated, with the zero-temperature approximation, the internal quantum efficiency in junctions involving 3D materials (metals) by the following [21]:

$$\eta_{int}^{3D} = \frac{N}{N_T} = \frac{1}{8q\phi_{Bo}} \cdot \frac{(h\nu - q\phi_{B0})^2}{h\nu} \qquad (4)$$

being $\phi_{B0}$ the Schottky barrier height (SBH) at zero bias, $h\nu = 1242/\lambda_0$[nm] the photon energy ($\lambda_0$ is the wavelength in vacuum condition) and q the electron charge. Very often a generic factor C (named quantum efficiency coefficient) replaces the factor $1/8q\phi_{B0}$ in order to achieve a better agreement between the theory and the experimental data. In order to achieve the expression (4) it is necessary to take $P(E) = (1 - \cos\vartheta)/2$, where $\vartheta$ is named carrier escape angle [20]. Elabd and Kosonocky in their work [21] outline also as the diminishing of the thickness of the metal causes an enhancement of the efficiency due to the increased emission probability P(E).

The 3D apex in the Elabd and Kosonocky equation (4) indicates the internal quantum efficiency for a metal-based junction, i.e. for a 3D material, but this equation fails to correctly describe the behaviour of a junction based on 2D materials [34,35] due to the different expressions to use for both the density of state D(E) and the emission probability P(E).

This issue has been discussed in detail for graphene [34,36]. Graphene has a band structure characterized by valence and conduction bands which touch in six points of the Brillouin zone. These points are termed Dirac points and represent the zero level of energy. In the graphic representation of the band diagram of a graphene/n-Si Schottky junction (fig. 2) one of these Dirac points is represented with a conical surface [37].

Unlike metals, Graphene shows a density-of-state function D(E) linearly dependent on the energy according to the formula: $D(E) = \frac{2|E|}{n\hbar^2 v_F^2}$ [38], where $\hbar$ is the reduced Plank constant and $v_F$ is the Fermi velocity. On the other hand, as discussed in Ref. [27], the emission probability P(E) simply can be taken equal to 1/2 because the graphene π orbital are always perpendicular to graphene/Si interface and therefore the momenta of the photo-excited carriers can only have two directions: towards the Si semiconductor or in the opposite direction [34].

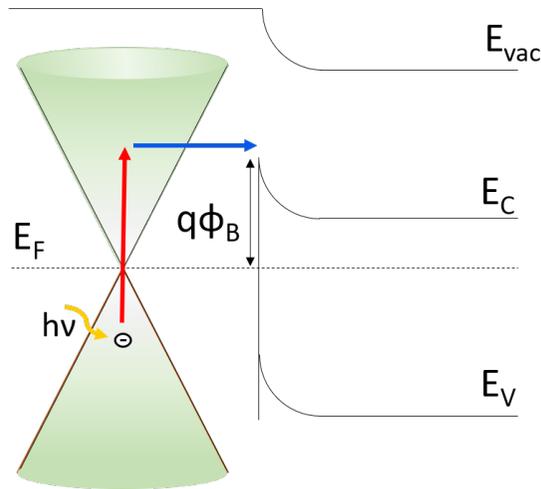

Figure 2. Energy band diagram of a graphene/n-Si Schottky junction with the conical representation of a Dirac point. $E_V$ ($E_C$): silicon valence (conduction) band; $E_F$: metal Fermi level; $q\Phi_B$: Schottky barrier.

Downstream of all these considerations and taking advantage of equations (2-3), the graphene quantum efficiency can be written:

$$\eta_{int}^{2D} = \frac{N}{N_T} = \frac{1}{2} \cdot \frac{(h\nu)^2 - (q\phi_{B0})^2}{(h\nu)^2} \qquad (5)$$

where the apex 2D indicates that the formula is referred to a bi-dimensional material [34,35].

From the plot of the (5) and (4) it is evident that the IPE effect is enhanced by using graphene material, as showed in the fig. 3 where the trend of the internal quantum efficiency versus the wavelength is reported for three different SBHs.

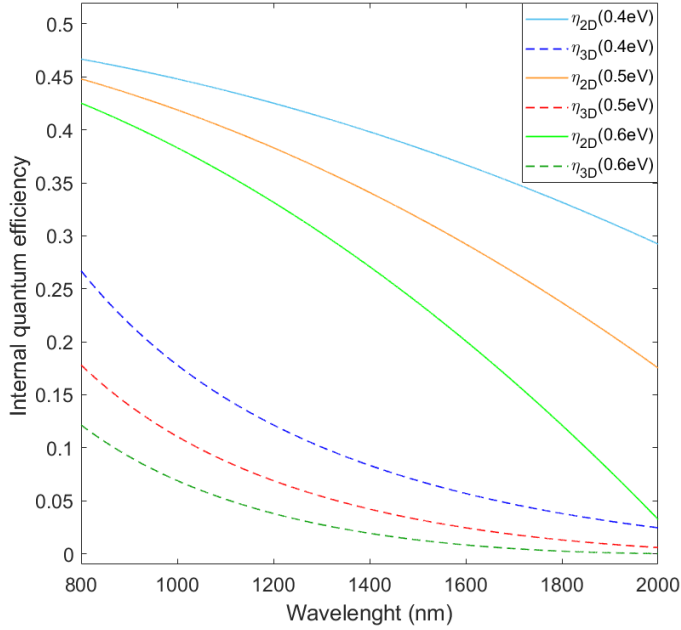

Figure 3. Comparison between $\eta_{int}^{3D}$ and $\eta_{int}^{2D}$ at vary wavelengths for three different value of Schottky barriers: 0.4, 0.5 and 0.6eV.

## 3. Schottky silicon photodetectors based on 2D materials

In last years, graphene has revolutionized the world of photonics and electronics thanks to its exceptional properties. Since its discovery, many researchers have concentrated their efforts on the possibility to integrate the graphene into optoelectronic devices. Notably, its zero direct bandgap make it very attractive for photodetection on a wide range from UV to IR. In particular, the demonstration of the graphene/silicon Schottky junction [39] has opened the path to realize more efficient NIR photodetectors exploiting the IPE.

In 2013, Amirmazlaghani et al. investigated a NIR PD based on exfoliated graphite on the top of a silicon substrate [34]. The Schottky barrier at the interface between the two materials resulted 0.44-0.47 eV and the ideality factor was 1.3-2.1. When a reverse bias of 16V was applied, the device exhibited a dark current of the order of μm and, under a 1.55 μm illumination, a maximum responsivity of 9.9 mAW[-1]. This value, higher than the one predicted by the equation (4), was explained by the authors as a consequence of the of the linear dispersion in graphene that requires a correction of the modified Fowler theory. By taking into account the two-dimensional nature of the graphene they derived the eq. (5) able to provide a better agreement with the experimental data. This issue was confirmed by Goykhman et al. who in 2016 demonstrated an increase in efficiency of 7% with respect to the values predicted by the eq. (4). The device investigated in [4] is a 5μm silicon waveguide covered by a layer of graphene. The plasmonic enhancement was obtained thanks to a film of Au on the top of the graphene. At 1V reverse bias the authors reported a responsivity of 85mAW[-1], that could grow up to 0.37AW[-1] at a reverse voltage of -3V. This happens thanks to an avalanche multiplication effect that unfortunately caused an abrupt increment of the dark current from 20nA to 3μA. Recently, Levy et al. [35] have proposed a phenomenological theory to explain the enhancement of internal photoemission in gold/graphene/silicon plasmonic structures.

In 2017 Casalino et al. realized vertically illuminated resonant cavity enhanced PDs exploiting the IPE through a CVD grown Single Layer Graphene (SLG) placed on top of a silicon substrate provided of a gold mirror on the back which acted as an optical cavity (fig.4a) [40]. This optical microcavity allowed to trap the radiation increasing the light round-trips in the cavity and enhancing the SLG optical absorption. A wavelength-dependent photoresponse was achieved (fig. 4b) with a peak value of the responsivity at 1.55µm of 20mAW$^{-1}$ at -10V applied voltage. At such bias the dark current was 147µA. The authors also evaluated the NEP and the bandwidth of the devices, that resulted 3.5 x 10$^{-10}$ WHz$^{-1}$ and 120MHz, respectively. Furthermore, it is worth noting that the Fabry-Perot cavity with a finesse of 5.4 determined a high spectral selectivity that could be easily tuned by changing the length of the resonant structure. The same author has devised another device, theoretically investigated in [41,42], where the SLG was situated in the centre of c-Si/a-Si:H optical cavity. The photodetection mechanism is based on IPE through the SLG/c-Si junction. The resonant structure, embedded between two high reflectivity dielectric mirrors, enabled an increased number of round-trips of the radiation that crossed multiple times the graphene layer strongly increasing its absorption. This not only provides a 100% maximum SLG absorption but also a responsivity and a finesse of 0.43 A/W and 172 in a correctly designed PD. Further, in this work the bandwidth and the noise of the device were discussed. In addition, a similar device taking advantage of a double silicon on insulator substrate working as a high-reflectivity mirror has been recently proposed and theoretically discussed [43].

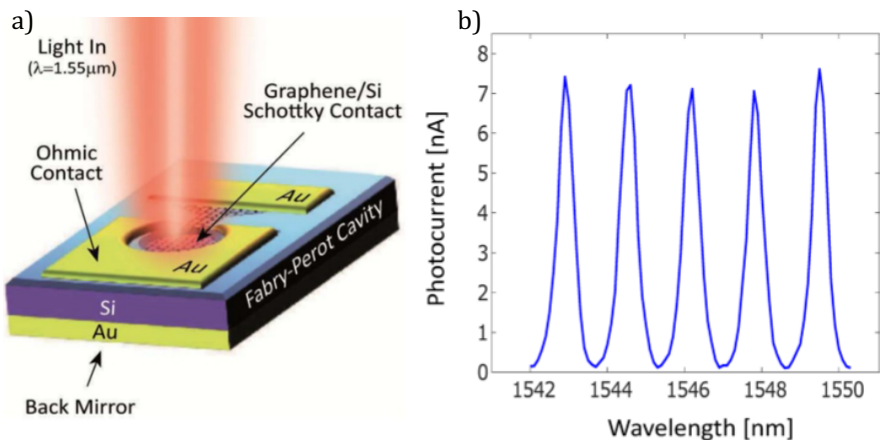

Figure 4. Sketch (a) and PDs spectral photoresponse (b) of the resonant cavity enhanced PDs investigated by Casalino et al. [40].

In 2016 Cheng et al. [44] demonstrated graphene short-wave NIR PDs with a very high responsivity of 83A/W at 1.55µm thanks to the combination of two different mechanisms that allow the improvement of the performances of their devices. Indeed, they overcome the problems of the low optical absorption and the short lifetime of the photogenerated charge carriers by exploiting plasmonic effects and a vertical built-in field at the graphene/silicon interface. As shown in fig. 5a the plasmonic effect, obtained thanks to gold nanoparticles array on the graphene channel, enables the trapping of photons resulting in a greater absorption (fig. 5b). Additionally, the vertical built-in potential induces a sort of carrier-trapping effect, by guiding the electrons away of the graphene and thus by generating holes with a consequent longer carrier lifetime. Indeed, the extension of the built- in field along all the large heterojunction produces a diminishing of the carrier recombination.

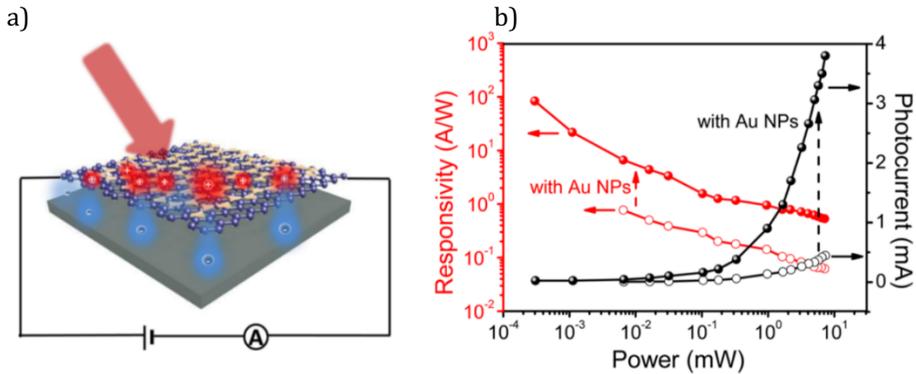

Figure 5. a) Schematic illustration of the graphene SWIR PD reported in [44]. In b) the comparison between the photoresponse of the devices with and without gold nanoparticles at vary illumination powers.

Recently, it has been proved that graphene/Si PDs based on the IPE can operate also at wavelengths greater than 1.55um. In [39] Casalino et al. reported the first demonstration of free-space vertically-illuminated PDs operating under a 2 μm radiation. Through an electrical analysis in a range of temperature from 280 to 315°C, they extracted the value of the SBH resulted to be 0.62eV at 300K. From the analysis it emerged a temperature dependence of the SBH which has been ascribed to the presence of defect at the interface between graphene and silicon. The proposed devices show at zero bias an internal responsivity of 10.3 mA/W, corresponding to an external one of 0.16 mA/W, accordingly to the theorical predictions.

In last years, there has been increasing interest in others 2D layered materials. In particular, TMDCs have emerged thanks to the attractive possibility to tune their bandgaps through the quantity of layers as well as their exceptional electronic and optical properties.

Molybdenum disulfite ($MoS_2$) is characterized by an indirect bandgap of about 1.3eV that increases up to 1.8eV and changes into a direct one in the monolayer.

In 2015, Wang et al. presented a $MoS_2$/Si heterojunction based on vertically standing layered configuration for the realization of ultrafast photodetectors [28]. The deposition of $MoS_2$ via sputtering allowed the growth of a policrystalline film with a vertical structure, from the p-silicon substrate up to the Ag electrode (fig. 6a), enabling the exploitation of the high in-plane mobility of the $MoS_2$. The electrical analysis of the junctions showed a potential barrier at the interface between the two materials of 0.33eV while the good quality of the junction was proved by an ideal factor of 1.83 and a rectification ratio of about 5000. The PD worked over a broadband spectrum, from visible to near infrared, with a maximum responsivity of 300mAW$^{-1}$ at 808nm (fig. 6b). The low dark current of the junction ensured a high detectivity up to 10$^{13}$Jones and a fast response of 2µs. Furthermore, the PD exhibited a photovoltaic behaviour by producing a photovoltage and a photocurrent of 210mV and 100µA, respectively, at zero bias.

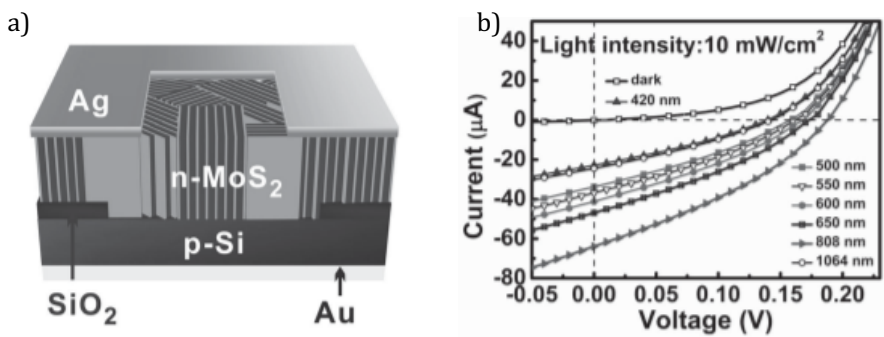

Figure 6. a) Schematic diagram of the MoS$_2$ /p-Si junction-based PDs realized in [28] and b) I-V characteristics of the device measured under different light illumination.

Subsequently, Kim et al. have proposed a similar PD based on a tungsten disulphite active layer [29]. Thanks to a bottom-up approach, by using a magnetron sputtering, they were able to grow vertical WS$_2$ layers onto a p-Si wafer at different temperatures fig. 7a. Through the X-Rays Diffraction (XRD) analysis they found a highly crystalline structure in the layers deposited at 400°C.

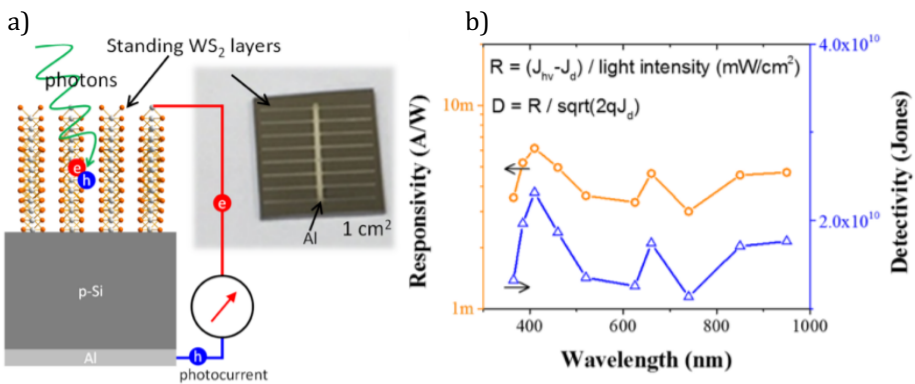

Figure 7. a) Sketch and photograph of the WS$_2$/p-Si based PD investigated by [29] et al. and b) Spectral photoresponse of the device.

The I-V curves demonstrated the formation of the heterojunction and the rectifying behaviour within ±2V where the rectification ratio was about 20000. The ideality factor and the dark saturation current I$_S$ were estimated to be 1.43 and 0.1 µA, respectively. The WS$_2$/Si junction exhibited a zero-bias photoresponse and an open-circuit voltage of 210mV together with a remarkable signal-to-noise ratio greater than 9000 for an incident radiation of 850nm. As shown in fig. 7b, the photoresponse of the device spanned the range of wavelengths from UV to NIR with peak responsivity values of 5-6mAW$^{-1}$ at 420, 680, 800 and 1000nm suggesting a an excitonic absorption of the WS$_2$. In addition, Kim et al. analysed the transient photocurrent at various wavelengths allowing to evaluate the photoresponse speed of the photodetectors that results to be about 1.1µs for a 10 kHz modulated signal, very higher than the conventional Si UV photodetectors. This impressive performance can be attributed to the large in-plane charge WS$_2$ carrier mobility.

Very interestingly, in 2019 Ahmad et al. reported a photodetector based on a WS$_2$ monolayer/Si junction [31]. The WS$_2$ monolayer was characterized by a lower bandgap with respect to the bulk material enabling higher responsivity of 10.46 mAW$^{-1}$ at 785nm. On the other hand, such configuration did not permit to take advantage of the in-plane conductivity of the absorber medium, resulting in a slower response of 186.7ms for a 20kHz modulated signal.

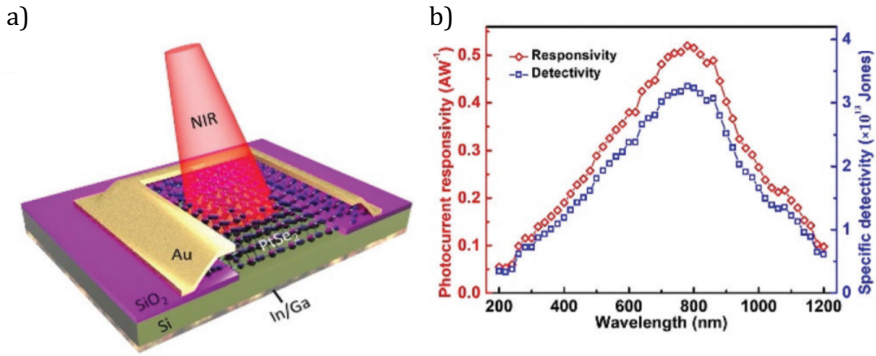

Figure 8. a) Schematic and spectral photoresponse (b) of the Pt/S$_2$/p-Si heterojunction PD reported by Xie et al. [31].

Another emerging 2D TMDC is the platinum diselenide (PtS$_2$). Its bandgap, ranging from zero in the monolayer to 1.2eV in the bulk, make it very promising for the NIR photodetection. Recently, Xie et al. investigated PDs exploiting a multilayered PtSe$_2$/silicon heterojunction (fig. 8a) [31]. In their work a thermally assisted conversion was used in order to have the *in situ* preparation of the PtSe$_2$ on the silicon substrate. Such technique permitted to realize interfaces with a small number of defects that would trap the photogenerated carriers. The XRD patterns displayed a policrystalline structure with nanometre-sized crystalline domains. The as-deposited 14.5nm-thick film, corresponding to about 17 layers of PtS$_2$, can be accordingly considered as a semimetal. The I-V curves confirmed the rectifying nature of the heterojunction in the range within ±5V and the ideality factor was estimated to be about 1.64. As displayed in fig. 8b the PD could operate in a wide spectrum ranging from 200 to 1550nm with a maximum responsivity of 520mAW$^{-1}$ at 808nm. This device showed also the capability to detect the telecommunication wavelengths of 1310 and 1550 nm with a responsivity of 33.25 and 0.57mAW$^{-1}$, respectively. Such results were attributed to the high NIR radiation optical absorption of the PtSe$_2$ layer. It is worth mentioning that these PDs showed a fast response, indeed, the rise time and fall time were 55.3 and 170.5 µs, respectively. The clean interfaces obtained thanks to the *in situ* preparation strongly influenced the performance of the device that exhibited a response speed comparable to the above mentioned works on TMDC PDs based on vertical structure.

**4. Conclusions**

In this chapter the physical principles of NIR Schottky PDs based on 2D materials have been elucidated and the main devices reported in literature have been discussed. In particular, PDs exploiting the IPE among 2D layered materials and silicon are deepened since to date they represent the most promising approach for the realization of high performances Si-based PDs. Devices discussed along this chapter have been summarized in Table 1 to allow an immediate comparison of their performance. It emerges that the low absorption coefficient of the graphene makes indispensable the use of structures enabling the light trapping for enhancing the light-matter interaction. Indeed, devices exploiting resonant cavities, waveguides and plasmonic effects result to have best performances in terms of responsivity. These structures show performance comparable with the well-established germanium technology adding the potentialities to detect wavelength longer than 1550nm. Although most of the Schottky PDs are based on graphene, more recently others 2D materials have stood out showing promising outcomes in the NIR spectrum.

| Ref. | Type | R | λ(μm) | $I_d$ | SBH (eV) | Config. |
|---|---|---|---|---|---|---|
| [34] | Exfoliated graphite/ p-Si | 9.9mAW$^{-1}$ at -16V | 1550 | ~2.4μA at -16V | 0.44-0.46 | Free-space |
| [4] | SLG/p-Si | 370mAW$^{-1}$ at −3V | 1550 | ~3μA at −3V | 0.34 | WG |
| [40] | SLG/p-Si | 20mAW$^{-1}$ at −10V | 1550 | ~147μA at −10V | 0.46 | Free-space |
| [41]* | SLG/p-Si | 0.43AW$^{-1}$ at 0V | 1550 | 561nA if SLG radius <15μm | 0.45 | Free-space |
| [44] | SLG/n-Si | 83AW$^{-1}$ | 1550 | ~0.1μA at -1.5V | 0.5 | Free-space |
| [35] | SLG/p-Si | 0.16mAW$^{-1}$ at 0V | 2000 | ~3μA at −6 V | 0.62 | Free-space |
| [28] | $MoS_2$/p-Si | 300mAW$^{-1}$ | 808 | - | 0.33 | Free-space |
| [29] | n-$WS_2$/p-Si | 5-6mAW$^{-1}$ | 420, 680, 800, 1000 | 0.1μA (saturation current) | - | Free-space |
| [30] | $WS_2$/n-Si | 10.46mAW$^{-1}$ | 785 | 0.1μA at −6V | - | Free-space |
| [31] | $PtS_2$/n-Si | 520mAW$^{-1}$/ 0.57mAW$^{-1}$ | 808/ 1550 | 1.1nA at -1V | - | Free-space |

Table 1: Comparison of the main electrical and optical parameters of the 2D materials/Si NIR PDs reported in this chapter.
\* Theoretical work

Thanks to the easy fabrication processes and the low cost of production, this new family of PDs represents a breakthrough, opening the way towards the commercial integration of silicon in photonics.

**References**

[1] Yole Dèvelop. [Internet]. 2020 Available from: http://www.yole.fr/Si_Photonics_Datacom_Sensing.aspx

[2] Soref, R. A., and J. P. Lorenzo: Single-crystal silicon: a new material for 1.3 and 1.6 µm integrated-optical components. Electronics Letters 21.21 1985; 953-954

[3] Soref, R.A.; Lorenzo, J.P. IEEE J. Quantum Electron. 1986, 22, 873–879

[4] Goykhman, I., Sassi, U., Desiatov, B., Mazurski, N., Milana, S., De Fazio, D., ... & Ferrari, A. C.: On-chip integrated, silicon–graphene plasmonic Schottky photodetector with high responsivity and avalanche photogain. Nano letters. 2016; 16.5: 3005-3013. DOI: 10.1021/acs.nanolett.5b05216

[5] Alavirad, M., Roy, L., & Berini, P.: Surface plasmon enhanced photodetectors based on internal photoemission. Journal of Photonics for Energy, 2016; 6.4: 042511

[6] Casalino, M.: Internal photoemission theory: Comments and theoretical limitations on the performance of near-infrared silicon Schottky photodetectors. IEEE Journal of Quantum Electronics, 2016; 52.4: 1-10

[7] Scales, C.; Berini, P.: Thin-Film Schottky Barrier Photodetector Models. IEEE J. Quantum Electron. 2010; 46, 633–643

[8] Elabd, H.; Villani, T.; Kosonocky, W.F.: Palladium-Silicide Schottky-Barrier IR-CCD for SWIR Applications at Intermediate Temperatures. IEEE ED Lett. 1982; 3, 89

[9] Elabd, H.; Villani, T.S.; Tower, J.R.: High density Schottky-barrier IRCCD sensors for SWIR applications at intermediated temperature. In Proceedings of the SPIE's Technical Symposium East, Arlington, VA, USA, 3–7 May 1982

[10] Kosonocky, W.F., Elabd, H., Erhardt, H.G., Shallcross, F.V., Villani, T., Meray, G., Cantella, M.J., Klein, J., Roberts, N. 64 ⇥ 128-Elements High-Performance PtSi IR-CCD Image Sensor. In Proceedings of the 1981 International Electron Devices Meeting, Washington, DC, USA, 7–9 December 1981

[11] Kosonocky, W.F., Elabd, H., Erhardt, H.G., Shallcross, F.V., Meray, G.M., Villani, T.S., Groppe, J.V., Miller, R., Frantz, V.L., Cantella, M.J., et al. Design and performance of 64 ⇥ 128-element PtSi Schottky-barrier IR-CCD focal plane array. In: Proceedings of the SPIE's Technical Symposium East, Arlington, VA, USA, 3–7 May 1982

[12] Wang, W. L., Winzenread, R., Nguyen, B., Murrin, J.J. High fill factor 512 x 512 PtSi focal plane array. In: Proceedings of the SPIE's 33rd Annual Technical Symposium, San Diego, CA, USA, 22 December 1989

[13] Crisci, T.; Moretti, L.; Casalino, M.: Theoretical Investigation of Responsivity/NEP Trade-off in NIR Graphene/Semiconductor Schottky Photodetectors Operating at Room Temperature. Applied Sciences, 2011; 11, 3398.

[14] Casalino, M.; Sirleto, L.; Moretti, L.; Della Corte, F.; Rendina, I.: Design of a silicon resonant cavity enhanced photodetector based on the internal photoemission effect at 1.55 µm. Journal of Optics A: Pure and applied optics, 2006; 8.10: 909

[15] Casalino, M.; Sirleto, L.; Moretti, L.; Gioffrè, M.; Coppola, G.; Rendina, I.: Silicon resonant cavity enhanced photodetector based on the internal photoemission effect at 1.55 micron: Fabrication and characterization. Appl. Phys. Lett. 2008; 92, 251104

[16] Zhu, S.; Chu, H.S.; Lo, G.Q.; Bai, P.; Kwong, D.L.: Waveguide-integrated near-infrared detector with self-assembled metal silicide nanoparticles embedded in a silicon p-n junction. Appl. Phys. Lett. 2012; 100, 61109

[17] Sobhani, A.; Knight, M.W.; Wang, Y.; Zheng, B.; King, N.S.; Brown, L.V.; Fang, Z.; Nordlander, P.; Halas, N.J.: Narrowband photodetection in the near-infrared with a plasmon-induced hot electron device. Nat. Commun. 2013; 4, 1643

[18] Knight, M.W.; Sobhani, H.; Nordlander, P.; Halas, N.J.: Photodetection with active optical antennas. Science 2011; 332, 702–704

[19] Berini,P.; Olivieri,A.; Chen,C.: Thin Au surface plasmon waveguide Schottky detectors on p-Si. Nanotechnology 2012; 23, 444011

[20] Akbari, A.; Tait, R.N.; Berini, P.: Surface plasmon waveguide Schottky detector. Opt. Express 2010; 18, 8505–8514

[21] Elabd, H.; Kosonocky, W.F.: Theory and measurements of photoresponse of thin film Pd2Si and PtSi Schottky-barrier detectors with optical cavity. RCA Rev. 1982; 43, 569

[22] Vickers, V.E.: Model of Schottky barrier hot-electron-mode photodetection. Appl. Opt. 1971; 10, 2190

[23] Wang, X.; Gan, X.: Graphene integrated photodetectors and opto-electronic devices—A review. Chin. Phys. B 2017; 26, 34201

[24] Koppens, F.H.L.; Mueller, T.; Avouris, P.; Ferrari, A.C.; Vitiello, M.S.; Polini, M.: Photodetectors based on graphene, other two-dimensional materials and hybrid systems. Nat. Nanotechnol. 2014; 9, 780–793

[25] Wang, Q. H., Kalantar-Zadeh, K., Kis, A., Coleman, J. N., & Strano, M. S.: Electronics and optoelectronics of two-dimensional transition metal dichalcogenides. Nature nanotechnology, 2012; 7.11: 699-712.

[26] Tang, Y., & Mak, K. F.: 2D materials for silicon photonics. Nature nanotechnology, 2017; 12.12: 1121-1122.

[27] Novoselov, K. S., Mishchenko, O. A., Carvalho, O. A., & Neto, A. C.: 2D materials and van der Waals heterostructures. Science; 2016, 353.6298.

[28] Wang, L., Jie, J., Shao, Z., Zhang, Q., Zhang, X., Wang, Y., ... & Lee, S. T: MoS2/Si heterojunction with vertically standing layered structure for ultrafast, high-detectivity, self-driven visible–near infrared photodetectors. Advanced Functional Materials, 2015; 25.19: 2910-2919

[29] Kim, H. S., Patel, M., Kim, J., & Jeong, M. S.: Growth of wafer-scale standing layers of WS2 for self-biased high-speed UV–Visible–NIR optoelectronic devices. ACS

applied materials & interfaces, 2018; 10.4: 3964-3974

[30] Ahmad, H., Rashid, H., Ismail, M. F., & Thambiratnam, K.: Fabrication and characterization of tungsten disulphide/silicon heterojunction photodetector for near infrared illumination. Optik, 2019; 185: 819-826

[31] Xie, C., Zeng, L., Zhang, Z., Tsang, Y. H., Luo, L., & Lee, J. H.: High-performance broadband heterojunction photodetectors based on multilayered PtSe 2 directly grown on a Si substrate. Nanoscale, 2018; 10.32: 15285-15293

[32] Fowler, R.H.: The analysis of photoelectric sensitivity curves for clean metals at various temperatures. Phys. Rev. 1931; 38, 45–56

[33] Cohen, J.; Vilms, J.; Archer, R.J.: Investigation of Semiconductor Schottky Barriers for Optical Detection and Cathodic Emission; Report No. 68-0651; Air Force Cambridge Research Labs: Force Base, OH, USA, 1968; p. 133

[34] Amirmazlaghani, M.; Raissi, F.; Habibpour, O.; Vukusic, J.; Stake, J.: Graphene-Si Schottky IR detector. IEEE J. Quant. Elect. 2013; 49, 2589

[35] Casalino, M., Russo, R., Russo, C., Ciajolo, A., Di Gennaro, E., Iodice, M., & Coppola, G.: Free-space Schottky graphene/silicon photodetectors operating at 2 μm. ACS Photonics, 2018; 5.11: 4577-4585

[36] Levy, U., Grajower, M., Goncalves, P. A. D., Mortensen, N. A., & Khurgin, J. B.: Plasmonic silicon Schottky photodetectors: The physics behind graphene enhanced internal photoemission. Apl Photonics, 2017; 2.2: 026103

[37] Goncalves, P.A.D.; Peres, N.M.R.: Electromagnetic properties of solids in a nutshell. In An Introduction to Graphene Plasmonics; World Scientific: Singapore, 2016; pp. 17–24. ISBN 978-981-4749-97-8

[38] Van Tuan, D. Electronic and Transport Properties of Graphene. In Charge and Spin Transport: in Disordered Graphene-Based Materials; Springer: Cham, Switzerland, 2016; p. 10

[39] Chen, C. C., Aykol, M., Chang, C. C., Levi, A. F. J., & Cronin, S. B.: Graphene-silicon Schottky diodes. Nano letters, 2011; 11.5: 1863-1867

[40] Casalino, M., Sassi, U., Goykhman, I., Eiden, A., Lidorikis, E., Milana, S., ... & Ferrari, A. C.: Vertically illuminated, resonant cavity enhanced, graphene–silicon Schottky photodetectors. ACS nano, 2017; 11.11: 10955-10963

[41] Casalino, M.: Design of resonant cavity-enhanced schottky graphene/silicon photodetectors at 1550 nm. Journal of Lightwave Technology, 2018; 36.9: 1766-1774

[42] Casalino, M., Crisci, T., Moretti, L., Gioffrè, M., Iodice, M., Coppola, G., ... & Morandi, V.: Silicon Meet Graphene for a New Family of Near-Infrared Resonant Cavity Enhanced Photodetectors. In: 2020 22nd International Conference on Transparent Optical Networks (ICTON); 19-23 July; Bari, Italy: IEEE; 2020. p. 1-4

[43] Casalino, M.: Theoretical Investigation of Near-Infrared Fabry–Pérot Microcavity Graphene/Silicon Schottky Photodetectors Based on Double Silicon on Insulator Substrates. Micromachines, 2020; 11.8: 708

[44] Chen, Z., Li, X., Wang, J., Tao, L., Long, M., Liang, S. J., ... & Xu, J. B.: Synergistic effects of plasmonics and electron trapping in graphene short-wave infrared photodetectors with ultrahigh responsivity. ACS nano, 2017; 11.1: 430-437.